\documentclass[floats,showpacs,pra,superscriptaddress,twocolumn]{revtex4}%
\usepackage{amsmath}
\usepackage{amsfonts}
\usepackage{amssymb}
\usepackage{graphicx}%
\providecommand{\U}[1]{\protect\rule{.1in}{.1in}}
\providecommand{\U}[1]{\protect\rule{.1in}{.1in}}
\providecommand{\U}[1]{\protect\rule{.1in}{.1in}}
\providecommand{\U}[1]{\protect\rule{.1in}{.1in}}
\providecommand{\U}[1]{\protect\rule{.1in}{.1in}}
\providecommand{\U}[1]{\protect\rule{.1in}{.1in}}
\providecommand{\U}[1]{\protect\rule{.1in}{.1in}}
\providecommand{\U}[1]{\protect\rule{.1in}{.1in}}
\providecommand{\U}[1]{\protect\rule{.1in}{.1in}}

\begin{document}
\title{Effective one-body dynamics in multiple-quantum NMR experiments}
\author{E. Rufeil Fiori}
\email{rufeil@famaf.unc.edu.ar}
\affiliation{Facultad de Matem\'{a}tica, Astronom{\'{\i}}a y F{\'{\i}}sica and Instituto de
F\'{\i}sica (CONICET), Universidad Nacional de C\'{o}rdoba, Ciudad
Universitaria, 5000, C\'{o}rdoba, Argentina.}
\author{C. M. S\'{a}nchez}
\affiliation{Facultad de Matem\'{a}tica, Astronom{\'{\i}}a y F{\'{\i}}sica and Instituto de
F\'{\i}sica (CONICET), Universidad Nacional de C\'{o}rdoba, Ciudad
Universitaria, 5000, C\'{o}rdoba, Argentina.}
\author{F. Y. Oliva}
\affiliation{INFIQC-Departamento de Fisicoqu\'{\i}mica, Facultad de Ciencias Qu\'{\i}micas,
Universidad Nacional de C\'{o}rdoba, Ciudad Universitaria, 5000, C\'{o}rdoba, Argentina.}
\author{H. M. Pastawski}
\affiliation{Facultad de Matem\'{a}tica, Astronom{\'{\i}}a y F{\'{\i}}sica and Instituto de
F\'{\i}sica (CONICET), Universidad Nacional de C\'{o}rdoba, Ciudad
Universitaria, 5000, C\'{o}rdoba, Argentina.}
\author{P. R. Levstein}
\email{patricia@famaf.unc.edu.ar}
\affiliation{Facultad de Matem\'{a}tica, Astronom{\'{\i}}a y F{\'{\i}}sica and Instituto de
F\'{\i}sica (CONICET), Universidad Nacional de C\'{o}rdoba, Ciudad
Universitaria, 5000, C\'{o}rdoba, Argentina.}

\begin{abstract}
A suitable NMR experiment in a one-dimensional dipolar coupled spin system
allows one to reduce the natural many-body dynamics into effective one-body
dynamics. We verify this in a polycrystalline sample of hydroxyapatite (HAp)
by monitoring the excitation of NMR many-body superposition states: the
multiple-quantum coherences. The observed effective one-dimensionality of HAp
relies on the quasi one-dimensional structure of the dipolar coupled network
that, as we show here, is dynamically enhanced by the quantum Zeno effect.
Decoherence is also probed through a Loschmidt echo experiment, where the time
reversal is implemented on the double-quantum Hamiltonian, $\mathcal{H}%
_{DQ}\propto I_{i}^{+}I_{j}^{+}+I_{i}^{-}I_{j}^{-}$. We contrast the
decoherence of adamantane, a standard three-dimensional system, with that of
HAp. While the first shows an abrupt Fermi-type decay, HAp presents a smooth
exponential law.

\end{abstract}
\startpage{1}
\maketitle

\section{Introduction}

The new developments in nanodevices \cite{Pett05, Luk07}, spintronics
\cite{AF07} and quantum information processing \cite{DiVic95} critically rely
on the control of quantum dynamics. This control is challenging because the
manipulation of quantum states \cite{Van07} is crucially limited by
decoherence \cite{Zur03, ZCP07}. In this sense, much can be learned from
nuclear magnetic resonance \cite{Cory97, Van04}, which offers the opportunity
to tailor the interactions, and thus the time scales, and to quantify
decoherence by implementing Loschmidt echoes \cite{JP01}.

The control of interaction anisotropy, e.g., the switch from a dipolar to an
$XY$ (planar) interaction, provides a tool for enhancing the transfer of
quantum information \cite{ADLP06, ADLP08}. In particular, the interactions can
be sequentially turned on and off to prune some branches in real space so that
an excitation is directed to a desired target through a specific pathway
\cite{Alv07}. By exploiting the mapping between spins and fermions, spin state
transfer in linear spin chains and rings coupled by $XY$ interaction was
proposed \cite{PUL96} and observed in liquid-state NMR \cite{MBS+97}.
Moreover, new suggestions that improve state transfer have been reported
\cite{B03, CDEL04, K07}. The\ structurally quasi-one-dimensional spin systems
of hydroxyapatite (HAp) and fluorapatite have been proposed as candidates for
implementing quantum information processing in solid-state NMR \cite{LGDYY}.
In these systems, universal control has been achieved by implementing
collective control together with suitable spin manipulation at the chain ends
\cite{CRC07, CRC207}.\emph{ }All these fine control attempts might be
frustrated by decoherence \cite{PLURH00, LUP98, KS06}. Thus, the dependence of
decoherence on nuclear spin network topology becomes an important issue.

In this work, we tailor the interactions in a one-dimensional (1d) dipolar
coupled spin system to transform its natural many-body dynamics into effective
one-body dynamics. The difference in dynamics is observed through the
excitation of NMR many-body superposition states: the multiple-quantum
coherences \cite{BMGP85}. Each $M$-quantum coherence ($M$-QC) collects all the
superpositions between two Zeeman states whose difference in total magnetic
moment is the integer $M$\textbf{. }$M$-QC intensities are tested in
solid-state NMR through phase codification techniques that allow one to follow
the superposition weights as they are being created \cite{Ern87}.

In a homogeneous one-dimensional chain of nuclear spin $1/2$, all spin sites
have the same energy and couplings. If the spins are coupled under
double-quantum interactions, $\mathcal{H}_{DQ}\propto I_{i}^{+}I_{j}^{+}%
+I_{i}^{-}I_{j}^{-}$, restricted to nearest-neighbors (NNs), analytical
methods give closed expressions for the intensities of the multiple-quantum
coherences \cite{DMF00}. Although $\mathcal{H}_{DQ}$ acting on a thermal
equilibrium state excites all even-order coherences, it can be proved that in
a one-dimensional system only zero- and second-order coherences are allowed
\cite{DMF00}. The results of this model are compared with numerical
calculations that include more realistic interactions and with NMR experiments
in a polycrystalline sample of HAp. HAp behaves as a quasi-one-dimensional
spin chain due to its dipolar coupled network structure \cite{CY93, CY96}. We
show that this anisotropy is further enhanced by a dynamical quantum Zeno
effect (QZE).

Decoherence is tested experimentally in HAp through a Loschmidt echo variant
\cite{JP01} based on $\mathcal{H}_{DQ}$ and its reversal. The same experiment
is performed in adamantane, a typical three-dimensional (3d) system, allowing
us to contrast the effect of the coupling network.

This paper is organized as follows. Section II discusses the multiple-quantum
coherences as well as the double-quantum Hamiltonian. Here, the theoretical
basis that allows one to obtain the effective one-body dynamics is summarized.
Section III describes the crystallographic and dynamical properties of HAp
which make it an effective one-dimensional system. Section IV describes the
experimental methods. Sections V and VI, respectively, present numerical and
experimental results for the $M$-QC dynamics. Section VII is devoted to the conclusions.

\section{Multiple-quantum coherence and effective one-body dynamics}

In a typical solid-state NMR experiment on a system of $N$ identical spins
$1/2$, the main interaction can be described by a dipolar Hamiltonian
truncated with respect to the dominant Zeeman interaction \cite{Sli92}:%
\begin{align}
\mathcal{H}_{ZZ}  &  =%
{\displaystyle\sum\limits_{i,j}}
\frac{d_{ij}}{2}\left(  2I_{i}^{z}I_{j}^{z}-I_{i}^{x}I_{j}^{x}-I_{i}^{y}%
I_{j}^{y}\right) \\
&  =%
{\displaystyle\sum\limits_{i,j}}
\frac{d_{ij}}{2}\left(  2I_{i}^{z}I_{j}^{z}-\frac{I_{i}^{+}I_{j}^{-}+I_{i}%
^{-}I_{j}^{+}}{2}\right)  ,
\end{align}
where $d_{ij}=(\gamma^{2}\hbar^{2}/(2r_{ij}^{3}))(3\cos^{2}(\theta_{ij})-1)$
are the dipolar couplings, with $\theta_{ij}$ as the angle between the
internuclear vector $\mathbf{r}_{ij}$ and the external magnetic field, and
$\gamma$ as the gyromagnetic ratio. $I_{i}^{z}$ are the $z$ components of the
spin operators defined by the direction of the static magnetic field, and
$I_{i}^{+}$and $I_{i}^{-}$ are the raising and lowering operators. In dipolar
coupled spin systems at high magnetic field, the off-diagonal elements of the
density matrix in the $z$ basis, i.e., the coherences $\rho_{rs}=\left\langle
r\right\vert \rho\left\vert s\right\rangle $, can be labeled by the difference
in the total magnetic quantum numbers between the states involved in the
transition, $M=m_{r}-m_{s}$, where $I^{z}\left\vert s\right\rangle
=m_{s}\left\vert s\right\rangle $, with $I^{z}=%
{\textstyle\sum_{i}}
I_{i}^{z}$. All the elements of the density matrix that connect two states
whose difference in total magnetic moment is $M$ contribute to the intensity
of an $M$-QC \cite{KS06}. Although only single-quantum coherences $(M=\pm1)$
are directly observed by NMR, phase codification techniques \cite{BMGP85}
allow one to obtain information on the multiple-quantum coherences.%

\begin{figure}
[h]
\begin{center}
\includegraphics[
height=1.0581in,
width=1.9113in
]%
{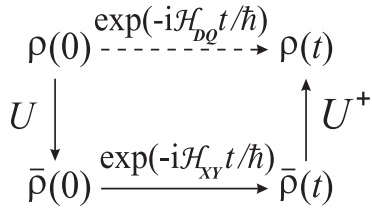}%
\caption{Pathways to generate multiple-quantum coherences from the initial
state $\rho(0)$. Experimentally one follows the dashed arrow. This is
equivalent, in a 1d system with nearest-neighbor interactions, to the
mathematical pathway indicated in solid arrows.}%
\label{Fig_esquema}%
\end{center}
\end{figure}

In order to create coherences from an initial thermal equilibrium state, a
Hamiltonian which does not commute with its density matrix is necessary. Both,
the dipolar Hamiltonian rotated to the $x$ axis ($\mathcal{H}_{XX}$) and the
double-quantum Hamiltonian ($\mathcal{H}_{DQ}$) fulfill this requirement and
are experimentally achievable:%
\begin{align}
\mathcal{H}_{XX}  &  =\exp(-\mathrm{i}\frac{\pi}{2}I_{{}}^{y})\mathcal{H}%
_{ZZ}\exp(\mathrm{i}\frac{\pi}{2}I_{{}}^{y})\\
&  =%
{\displaystyle\sum\limits_{i,j}}
\frac{d_{ij}}{2}\left(  2I_{i}^{x}I_{j}^{x}-I_{i}^{y}I_{j}^{y}-I_{i}^{z}%
I_{j}^{z}\right)  ,\label{Eq_Hxx_x}\\
\mathcal{H}_{DQ}  &  =%
{\displaystyle\sum\limits_{i,j}}
\frac{d_{ij}}{2}\left(  I_{i}^{x}I_{j}^{x}-I_{i}^{y}I_{j}^{y}\right) \\
&  =%
{\displaystyle\sum\limits_{i,j}}
\frac{d_{ij}}{4}\left(  I_{i}^{+}I_{j}^{+}+I_{i}^{-}I_{j}^{-}\right)  .
\label{Eq_Hdq_mas}%
\end{align}

In the special case of NN interactions, $\mathcal{H}_{DQ}$ is unitary similar
to the $XY$ Hamiltonian, $\mathcal{H}_{XY}\propto I_{i}^{+}I_{j}^{-}+I_{i}%
^{-}I_{j}^{+}$. Consequently, $\mathcal{H}_{DQ}$ can simulate the
$\mathcal{H}_{XY}$ dynamics after the corresponding transformation of the
initial state. Although this relation between $\mathcal{H}_{DQ}$ and
$\mathcal{H}_{XY}$ is valid in one, two and three-dimensions \cite{CRC07,
PMCC06}, we focus on one-dimensional systems, for which\ closed analytical
results are available. Here, we summarize the successive transformations,
developed by Doronin et al. \cite{DMF00}, that enable this mapping. First, one
applies the unitary transformation%
\begin{equation}
U=\exp(-i\pi I_{2}^{x})\exp(-i\pi I_{4}^{x})...\exp(-i\pi I_{2n}^{x})...,
\label{Eq_U}%
\end{equation}
to $\mathcal{H}_{DQ}$. This is a composition of $\pi$ pulses which rotate
even-numbered spins $180%
{{}^o}%
$ about the $x$ axis. As a result, the transformed Hamiltonian is%
\begin{equation}
\mathcal{H}_{XY}=U\mathcal{H}_{DQ}U^{\dagger}=%
{\displaystyle\sum\limits_{i}}
\frac{d_{i,i+1}}{4}\left(  I_{i}^{+}I_{i+1}^{-}+I_{i}^{-}I_{i+1}^{+}\right)  .
\label{Eq_Hxy}%
\end{equation}
The same transformation must be applied to the initial state. For the thermal
equilibrium state, in the high-field and high-temperature limit, we only
consider the main deviation of the density matrix from the identity, which is
the experimentally observable part, i.e., $\rho\left(  0\right)  =%
{\textstyle\sum\nolimits_{i}}
I_{i}^{z}$. This leads to%
\begin{equation}
\overline{\rho}\left(  0\right)  =U\rho\left(  0\right)  U^{\dagger}=%
{\displaystyle\sum\limits_{i}}
\left(  -1\right)  ^{i-1}I_{i}^{z}. \label{Eq_rho_techo}%
\end{equation}
Then, as shown schematically in Fig. \ref{Fig_esquema}, the dynamics of an
initial state $\rho\left(  0\right)  $ under $\mathcal{H}_{DQ}$\ is reduced to
the dynamics of $\overline{\rho}\left(  0\right)  $ under $\mathcal{H}_{XY}$
which, in turn, maps to a non-interacting fermion system \cite{LSM61, FR99,
DPL04}. The dynamics of this fermionic system has a closed analytical solution
when the interaction is homogeneous, $d_{i,i+1}=d,$ $\forall i$. Transforming
back to the double-quantum dynamics, a closed expression for the density
matrix $\rho\left(  t\right)  $ can be obtained. The intensities $J_{M}$ of
the $M$-QC are calculated as
\begin{equation}
J_{M}\left(  t\right)  =\mathrm{Tr}\left\{  \rho_{M}\left(  t\right)
\rho_{-M}\left(  t\right)  \right\}  , \label{Eq_Jm}%
\end{equation}
where
\begin{equation}
\rho_{M}\left(  t\right)  =\sum_{r,s}{}^{^{\prime}}\rho_{rs}\left(  t\right)
,
\end{equation}
where $%
{\textstyle\sum^{\prime}}
$ restricts the sum to $m_{r}-m_{s}=M$. Thus, $\rho_{M}$ collects all the
contributions to $\rho$ due to coherences of order $M$, and $\rho\left(
t\right)  =%
{\textstyle\sum\nolimits_{M}}
\rho_{M}\left(  t\right)  $.\ Then, the $J_{M}\left(  t\right)  $, in the
normalized form $\sum_{M}J_{\pm M}=1$, result in%
\begin{align}
J_{0}\left(  t\right)   &  =\frac{1}{N}%
{\textstyle\sum\limits_{n}}
\cos^{2}\left(  4\mathrm{d}t/\hbar\cos\left(  \frac{\pi n}{N+1}\right)
\right)  ,\\
J_{\pm2}\left(  t\right)   &  =\frac{1}{2N}%
{\textstyle\sum\limits_{n}}
\sin^{2}\left(  4\mathrm{d}t/\hbar\cos\left(  \frac{\pi n}{N+1}\right)
\right)  , \label{Eq_J2analitica}%
\end{align}
with $n=1,...,N$. This shows that only $Z$-QC and $2$-QC are allowed. All
other even-orders can not be created. Even though a closed analytical solution
is not possible in a NN inhomogeneous case, it was shown that only zero- and
second-order coherences are excited \cite{DF05}, as what occurs in the
homogeneous chain.

Finally, the evolution of particular initial conditions \cite{CRC07} under 1d
nearest-neighbor double-quantum interactions reduces to that of
non-interacting (\textquotedblleft\textit{one-body}\textquotedblright)
spinless fermions. This one-body dynamics manifests through the presence of
only 2 orders of coherence ($Z$-QC and $2$-QC). We test\ this in Secs. V and
VI by performing numerical simulations and multiple-quantum NMR experiments of
the dynamics under $\mathcal{H}_{DQ}$. This is contrasted with the irreducible
many-body dipolar dynamics under $\mathcal{H}_{XX}$.

\section{Dynamical enhancement of the one-dimensionality by the quantum Zeno
effect}

We perform NMR experiments in a physical system that behaves as a
one-dimensional spin $1/2$ chain. The system is a polycrystalline sample of
hexagonal hydroxyapatite, $\mathrm{Ca}_{5}\left(  \mathrm{PO}_{4}\right)
_{3}\mathrm{OH,}$ with space group P63/m. Due to the difference in resonance
frequencies of the various spin nuclei, the experimental setup allows taking
account of only the spin degrees of freedom of the $^{1}$H nuclei. The
hydrogen spins of this sample are ordered as linear chains in the $c$
direction of a hexagonal arrangement ($a=b,c$) \cite{CY96}. A central chain is
surrounded by six neighboring chains at a distance of $r_{\mathrm{x}%
}=9.42\ \mathring{A},\ $($r_{\mathrm{x}}=a$). The closest distance between
protons within a chain is $r_{\mathrm{in}}=3.44\ \mathring{A}$%
,\ ($r_{\mathrm{in}}=c/2$). In solid-state NMR the strongest interaction is
the dipolar one. Because of the dependence of the dipolar couplings on the
spin distance, the ratio between the in-chain, $d_{\mathrm{in}}$, and the
cross-chain, $d_{\mathrm{x}}$, dipolar couplings for the orientation that
maximizes the in-chain coupling is%
\begin{equation}
\frac{d_{\mathrm{in}}}{d_{\mathrm{x}}}=2\left(  \frac{r_{\mathrm{x}}%
}{r_{\mathrm{in}}}\right)  ^{3}\approx2\times20.
\end{equation}
As we work with a polycrystal, we calculate for each chain orientation the
ratio of the local second moment due to in-chain interactions,
$M_{2,\mathrm{in}}$, to the local second moment due to the six neighboring
chains, $M_{2,\mathrm{x}}$. Then, by taking the average over solid angle, we
obtain%
\begin{equation}
\sqrt{\left\langle \frac{M_{2,\mathrm{in}}}{M_{2,\mathrm{x}}}\right\rangle
}=\left\langle f(\theta,\phi)\right\rangle \left(  \frac{r_{\mathrm{x}}%
}{r_{\mathrm{in}}}\right)  ^{3}\approx1.5\times20.
\end{equation}
where $f\left(  \theta,\phi\right)  $ is the angular function that takes into
account the angular dependence of the dipolar interaction and the relative
orientation of the internuclear vectors with respect to the external magnetic field.

There is a dynamical effect that further enhances the difference between these
two couplings. The characteristic time for a flip-flop process within the
chain is clearly%
\begin{equation}
\tau_{\mathrm{in}}\approx\frac{\hbar}{d_{\mathrm{in}}}.
\end{equation}
However, the characteristic rate of a flip-flop due to the weak cross-chain
couplings should be estimated invoking the Fermi golden rule that yields
\cite{RP06}%
\begin{equation}
1/\tau_{\mathrm{x}}\approx\frac{1}{\hbar}d_{\mathrm{x}}^{2}\frac
{1}{d_{\mathrm{in}}}, \label{Eq_Zeno-Tx}%
\end{equation}
and not $d_{\mathrm{x}}/\hbar$ as one might first guess. This is because the
strong in-chain dynamics leads to an uncertainty of the final state over a
wide excitation spectrum. Then, we have%
\begin{equation}
\frac{\tau_{\mathrm{in}}}{\tau_{\mathrm{x}}}\approx\left(  \frac
{d_{\mathrm{x}}}{d_{\mathrm{in}}}\right)  ^{2}\approx\left(  \frac
{r_{\mathrm{in}}}{r_{\mathrm{x}}}\right)  ^{6}\approx\frac{1}{400}.
\label{Eq_Zeno-rate}%
\end{equation}
Equation (\ref{Eq_Zeno-Tx}) states that fast in-chain dynamics makes already
slow cross-chain dynamics even slower. This is a form of the QZE, which states
that quantum dynamics is slowed down by a frequent measurement process
\cite{MS77}. Spin-diffusion experiments in low-dimensional crystals showed an
unexpected dimensional cross-over as a function of a structural parameter
\cite{LPC91}. This cross-over was described as a QZE where the internal
degrees of freedom act as measurement apparatus \cite{PU98}. The concept that
the measurement is played by an interaction with another quantum object, or
simply another degree of freedom of the subsystem investigated, was
independently and fully formalized by recasting it in terms of an adiabatic
theorem in Ref. \cite{FP02}. It can even lead to a freeze of the spin swap
dynamics as observed in a cross-polarization experiment \cite{ADLP06}. In the
present context, Eq. (\ref{Eq_Zeno-rate}) reinforces the effective
one-dimensional behavior of HAp.

\section{NMR experimental setup}

The experiments were performed using a Bruker Avance II spectrometer operating
at a $^{1}H$ resonance frequency of 300.13 MHz. We used a cross polarization
magic angle spinning probe working in static conditions at room temperature
with a 4mm outer diameter rotor.

\begin{figure}
[h]
\begin{center}
\includegraphics[
trim=-0.030725in 0.046134in 0.030725in -0.046134in,
height=3.1894in,
width=2.4907in
]%
{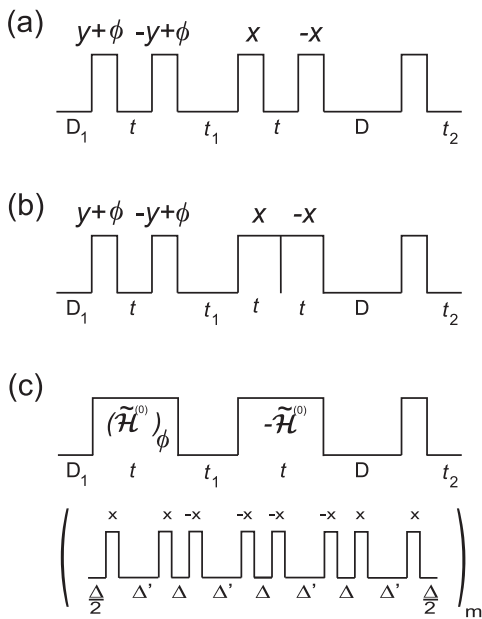}%
\caption{(a) Selective two-pulse sequence to generate $M$-QC under an average
double-quantum Hamiltonian $\mathcal{H}_{DQ}$ based on rotations of
$\mathcal{H}_{ZZ}$ (free evolution). It detects $2$-QC $(2+4n)$, or $4$-QC
$(4+8n)$ intensities by applying the appropriate phase cyclings (see text).
All pulses are of $\pi/2$. (b) Sequence to generate $M$-QC under a rotated
dipolar Hamiltonian $\mathcal{H}_{XX}$. (c) Pulse sequence to generate $M$-QC
under a $\mathcal{H}_{DQ}$ based on $m$ repetitions the eight $\pi/2$-pulse
pattern displayed in parenthesis. In (b) and (c) an free induction decay was
recorded for each value of $t$ and $\phi.$ The highest coherence order
detected, $n_{max}=8$, is governed by a digital phase shift increment, with
$\Delta\phi=\pi/n_{max}$.}%
\label{Fig_pulsos}%
\end{center}
\end{figure}

The characterization of the dynamics of the multiple-quantum coherences was
performed using the pulse sequences shown in Fig. \ref{Fig_pulsos}. The
different orders of coherence excited under $\mathcal{H}_{DQ}$ were generated
using the two-pulse sequence shown in Fig. \ref{Fig_pulsos}(a) \cite{VDPOB05,
AMVVV06}. With this sequence, $\mathcal{H}_{DQ}$ is built after a minimum
number of scans $N_{s}$, with $N_{s}=2M_{\mathrm{des}}$, where
$M_{\mathrm{des}}$ is the order of coherence one desires to detect indirectly.
Thus, in order to measure $2$-QC, a minimum of four scans must be added. To
get a better signal-to-noise ratio, the total number of scans must be a
multiple of $N_{s}$. Therefore, the evolution of $M_{\mathrm{des}}$-QC under
$\mathcal{H}_{DQ}$ is built after $N_{s}$ scans by adding signals with
different phases $\phi$. In particular, one uses $\phi=0,$ $\pi/2,$ $\pi$ and
$3\pi/2$ for filtering the $2$-QC and $\phi=0,$ $\pi/4,$ $\pi/2,$ $3\pi/4,$
$\pi,$ $5\pi/4,$ $3\pi/2$ and $7\pi/4$ for filtering the $4$-QC. In both
cases, the phase of the reading pulse was alternated between $0$ and $\pi$ to
keep only the orders of coherences\ $M_{\mathrm{des}}\pm nN_{s}$, with
$n=0,1,2,...$ \cite{Ern87, MP87}.%

In order to encode $M$-QC orders during the evolution under $\mathcal{H}_{XX}$
[Eq. (\ref{Eq_Hxx_x})], we used the sequence shown in Fig. \ref{Fig_pulsos}%
(b), which is a modified version of the sequence reported in Ref.
\cite{CLBCR05}. Here, the highest coherence order detected, $n_{max}=8$, is
governed by the phase shift increment $\Delta\phi=\pi/n_{max}$.

In these sequences, the recorded free induction decays were the sum of $64$
scans. The recycling time, $D_{1}=3\ s$, was chosen to be longer than five
times the spin-lattice relaxation time $T_{1}\approx500\ ms$. The $\pi/2$
pulse length was $2.74\ \mu s$. The preparation times $t$, i.e., the periods
evolving under the desired effective Hamiltonian, were varied from $1$ to
$200\ \mu s.$ The free evolution time $t_{1}=0.5\ \mu s$ was negligible. After
the mixing time and before the $\pi/2$ reading pulse, a delay $D=2\ ms$ was
used to allow the transverse magnetization to decay. The detected signal was
normalized to a reference FID obtained by the application of a $\pi/2$ pulse
with the same number of scans.

The two-pulse sequence used to generate $\mathcal{H}_{DQ}$ was chosen because
the fast growth of the $2$-QC intensity is not captured with the eight-pulse
sequence shown in Fig. \ref{Fig_pulsos}(c) \cite{BMGP85}. The last only
captures a few data points in the time range of interest because of the
minimum time of $\sim60\ \mu s$ required to accommodate the eight pulses of
the basic unit. However, the eight-pulse sequence was applied to implement a
\textquotedblleft Loschmidt echo\textquotedblright\ experiment, that is, to
generate $\mathcal{H}_{DQ}$ and then $-\mathcal{H}_{DQ}$, by using $\phi=0$.
We use this echo to give a measure of decoherence rates. In order to compare
this decoherence rate in HAp with a widely known system, we performed the echo
experiments in adamantane. Adamantane is a 3d molecular crystal with only
intermolecular dipolar interactions \cite{BP86} (the intramolecular
interactions cancel out due to rapid molecular rotations). In the Loschmidt
echo experiments, the preparation time was varied from $60\ $to $1400\ \mu s$
and the $\pi/2$ pulse length was $2.34\ \mu s$ for HAp and $2.20\ \mu s$ for adamantane.

The experiments were carried out in a polycrystalline sample of hydroxyapatite
synthesized by a modification of the biomimetic method reported by Zhang et
al. \cite{ZLX05}, while a commercial polycrystalline sample of adamantane was
used as provided.

\section{Numerical results: multiple-quantum dynamics}

The $M$-QC intensities were numerically simulated using an ensemble average of
the evolution of each Zeeman state. The total magnetization was calculated as
a function of preparation time $t$ and as a function of the $M$-QC
codification phase $\phi$. This was obtained by evolving each initial state
under $\mathcal{H}$ during $t$ and then under $-\mathcal{H}_{\phi}$, where
$\mathcal{H}_{\phi}=\exp(-\mathrm{i}\phi I^{z})\mathcal{H}\exp(\mathrm{i}\phi
I^{z})$. Finally, a fast Fourier transform on $\phi$ was applied to the
magnetization to obtain the $M$-QC intensities $J_{M}\left(  t\right)  $
\cite{BMGP85}.

An alternative method to obtain $J_{M}\left(  t\right)  $, which makes use of
Eq. (\ref{Eq_Jm}), was used. In this case, the $\rho_{M}$ are obtained from
the elements of the density matrix calculated for each Zeeman state. Although
this method is time consuming, it clearly shows how the different coherences
contribute to the intensity of a given order.

This second method allows us to draw some conclusions about the unitary
transformations schematized in Fig. \ref{Fig_esquema}. Even when
$\mathcal{H}_{DQ}$ is unitary similar to\ $\mathcal{H}_{XY}$, an arbitrary
initial condition under $\mathcal{H}_{DQ}$ does not necessarily yield only 2
orders of coherences. In a chain with NN $XY$ interaction, any excitation
remains in the same subspace, i.e., only zero-order coherences appear.
However, the transformed initial thermal equilibrium condition $\overline
{\rho}\left(  0\right)  $ [Eq. (\ref{Eq_rho_techo})] imposes a further
restriction in the accessible Hilbert space in which this condition can evolve
under $XY$ interaction. In this case, only a portion of the $ZQ$-subspace can
be reached. It is because of this restriction that, after transforming back to
the double-quantum dynamics $\rho(t)$, only zero- and second-order coherences
are excited.

\begin{figure}
[h]
\begin{center}
\includegraphics[
trim=-0.185268in 0.000000in -0.185268in 0.000000in,
height=2.5681in,
width=3.1382in
]%
{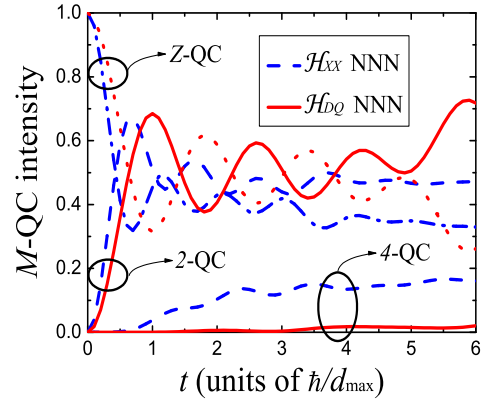}%
\caption{(Color online) Numerical simulations of the dynamics of the $Z$-QC$,$
$2$-QC and $4$-QC intensities of HAp under $\mathcal{H}_{XX}$ (dashed line for
$2$-QC and $4$-QC, and dash-dotted line for $Z$-QC) and $\mathcal{H}_{DQ}$
(solid line for $2$-QC and $4$-QC, and dotted line for $Z$-QC) in a\ $10$ spin
chain with nnn interaction for the chain orientation that maximizes the
coupling, $d_{\max}/\hbar=2\pi\times2.95$ kHz. The mesoscopic echo appears at
$6$ $\hbar/d_{\max}$.}%
\label{Fig_numerics_todas}%
\end{center}
\end{figure}

\begin{figure}
[h]
\begin{center}
\includegraphics[
trim=-0.128579in 0.000000in -0.128580in 0.000000in,
height=2.5036in,
width=3.1595in
]%
{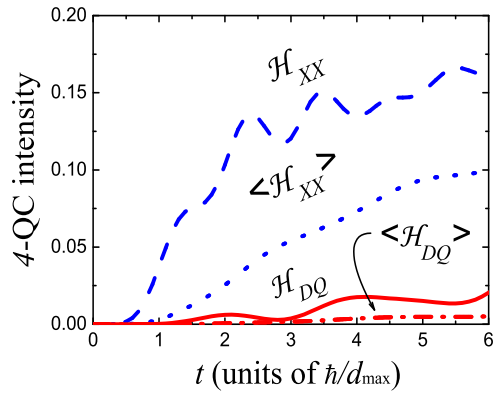}%
\caption{(Color online) Numerical simulations of the dynamics of the $4$-QC
intensities under $\mathcal{H}_{XX}$ and $\mathcal{H}_{DQ}$ for a\ $10$ spin
chain of HAp with nnn interactions. Dashed line corresponds to $\mathcal{H}%
_{XX}$ and solid line to $\mathcal{H}_{DQ},$ both at the orientation of
maximum dipolar coupling, $d_{\max}/\hbar=2\pi\times2.95$ kHz. The powder
average is shown with a dotted line for $<\mathcal{H}_{XX}>$ and with
dash-dotted line for $<\mathcal{H}_{DQ}>$.}%
\label{Fig_numerics}%
\end{center}
\end{figure}

In order to obtain the dynamics of $J_{M}\left(  t\right)  $ under
$\mathcal{H}_{DQ}$ and contrast this with that under\textbf{ }$\mathcal{H}%
_{XX}$, we used the first method described above. Since the effective
Hamiltonians $\mathcal{H}_{DQ}$ and $\mathcal{H}_{XX}$ are built up
experimentally from the natural dipolar interaction, which decays with
$1/r^{3}$, it becomes important to take into account the next-nearest-neighbor
(NNN) interaction in the simulations. In a chain, the values of the NNN
couplings are $1/8$ of the NN ones. The simulated dynamics of the $Z$-QC,
$2$-QC and $4$-QC intensities under $\mathcal{H}_{XX}$ and $\mathcal{H}_{DQ}$
is shown in Figs. \ref{Fig_numerics_todas} and \ref{Fig_numerics} for an
$N=10$ spin chain starting at thermal equilibrium. Preliminary experimental
results in polycrystalline HAp showed that there were no detectable $M$-QC
intensities after $200$ $\mu s$. Consequently, we do not need simulations for
longer times, but we have to take a large enough number of spins to avoid
distortions of the dynamics due to reflections at the chain ends. To verify
this, we calculated the earliest time at which the mesoscopic echo, i.e., the
revival that appears because of the finite nature of the system
\cite{PLU95,PUL96}, occurs. This is ensured by using a single crystal at
orientation $\theta_{ij}=0$, leading to the maximum dipolar coupling $d_{\max
}$, which for HAp is $d_{\max}=2\pi\hbar\times(2.95~\mathrm{kHz})$. Any other
orientation will just stretch the time scale of this curve, delaying the
occurrence of the mesoscopic echo. As shown in Fig. \ref{Fig_numerics_todas},
for ten spins the mesoscopic echo appears at $6$ $\hbar/d_{\max}\approx325$
$\mu s$. It is important to emphasize that by varying slightly the number of
spins, the dynamics changes only in the neighborhood of the mesoscopic echo,
remaining unaffected before $3.7$ $\hbar/d_{\max}\approx200$ $\mu s$.%

%

In Fig. \ref{Fig_numerics}, the $4$-QC intensity dynamics in a ten spin chain
of HAp with NNN interactions under $\mathcal{H}_{DQ}$ and $\mathcal{H}_{XX}$
is displayed. In each Hamiltonian evolution, a single orientation of the chain
and a powder average (the integral over solid angle of the orientation
dependent dynamics) were calculated. Notice that the observable we are using
to check the effective one-body dynamics is robust under orientation average,
i.e., the non-excitability of the $4$-QC occurs for every orientation of the
chains in a polycrystalline sample, maintaining its null intensity. As can be
seen, if one includes the NNN interactions in the chain, the intensity of the
$4$-QC under $\mathcal{H}_{DQ}$ is not strictly zero. However, this intensity
will not be observed under the typical conditions of an NMR experiment. In
contrast, the intensity of a $4$-QC under $\mathcal{H}_{XX}$ might be observable.

The inclusion of an extra interaction, in this case the NNN interaction,
breaks the mapping to non-interacting fermions. Consequently, the system
evolution is no longer restricted to only $Z$-QC and $2$-QC. However, as it is
clearly shown in Fig. \ref{Fig_numerics}, the $4$-QC\ under $\mathcal{H}_{DQ}$
is 1 order of magnitude smaller than the $4$-QC excited by $\mathcal{H}_{XX}$.
This means that $\mathcal{H}_{DQ}$ still keeps the main dynamics between
$Z$-QC and $2$-QC. Hence, one can infer that the effective one-body dynamics
is preserved as a good approximation.

\section{Experimental results: multiple-quantum dynamics and decoherence}

\begin{figure}
[h]
\begin{center}
\includegraphics[
height=2.5165in,
width=3.1668in
]%
{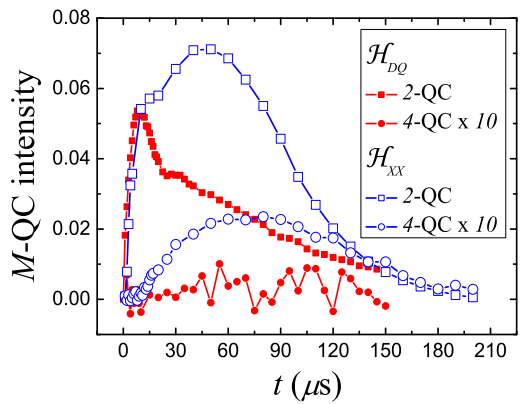}%
\caption{(Color online) Dynamics of $2$-QC and $4$-QC intensities under
$\mathcal{H}_{DQ}$ and $\mathcal{H}_{XX}$ in HAp implemented with the pulse
sequence of Fig. \ref{Fig_pulsos} (a) and (b), respectively. Notice that the
normalized intensities of the $4$-QC are 10 times enlarged.}%
\label{Fig_experiments_bu}%
\end{center}
\end{figure}

The pulse sequences shown in Figs. \ref{Fig_pulsos}(a) and \ref{Fig_pulsos}(b)
were used to generate $M$-QC under the effective Hamiltonians $\mathcal{H}%
_{DQ}$ and $\mathcal{H}_{XX}$, respectively, from a thermal equilibrium state.
Figure \ref{Fig_experiments_bu} displays the $2$-QC and $4$-QC intensities as
functions of the preparation time $t$. There, the $4$-QC has been multiplied
by a factor of $10$ because of its small intensity as compared with the
$2$-QC. While the $4$-QC intensity under $\mathcal{H}_{XX}$ is well above the
noise level, being evident its growth and decay, the intensity of the $4$-QC
under $\mathcal{H}_{DQ}$ remains at the noise level.%

In our particular 1d system, the essential difference between the ideal
$\mathcal{H}_{XX}$ and $\mathcal{H}_{DQ}$ is that the first allows for the
development of many orders of coherence, while the second allows only 2. Since
higher orders of coherences decay at higher rates \cite{KS04, SPL07}, we
expect a faster decoherence in the case of $\mathcal{H}_{XX}$. However, we
should also assess the precision of the experimental sequences used to
generate these Hamiltonians. In this context, one should remember that our
implementation of a multiple-quantum experiment under $\mathcal{H}_{XX}$
includes a dipolar Hamiltonian reversal. This involves a further truncation of
the dipolar Hamiltonian with respect to the rf Zeeman interaction during the
long rf pulse \cite{Sli92}. This produces additional decoherence because the
truncated non-secular terms, whose magnitudes depend on the rf power, are not
reversed \cite{LUP98}. The pulse sequences to generate $\mathcal{H}_{DQ}$ may
also have some limitations. It is known that the eight-pulse sequence produces
a much better average Hamiltonian than the two-pulse one, especially for long
preparation times \cite{BMGP85, Mun88}. This is because the two-pulse sequence
plotted in Fig. \ref{Fig_pulsos}(a) does not average out the chemical shift
nor cancels out rf inhomogeneities as the eight-pulse sequence does. For
example, if we compare the $2$-QC intensities in HAp for the eight- and
two-pulse sequences, both of them show exponential decay. However, the
characteristic time of the first is $\tau_{8p}\approx210\ \mu s$, while that
of the second is $\tau_{2p}\approx65\ \mu s$, i.e., it is three times faster.
The rapid decay of the $2$-QC intensity with the two-pulse sequence explains
the early occurrence of the maximum $(\approx15\ \mu s)$ in the evolution of
$2$-QC (see Fig. \ref{Fig_experiments_bu}) as compared with the theoretical
estimation of $\hbar/d_{\max}\approx50\ \mu s$ in Fig.
\ref{Fig_numerics_todas}. Because of this, the decay of the $2$-QC for the
two-pulse sequence is not a reliable quantifier of the decoherence of the system.

\begin{figure}
[h]
\begin{center}
\includegraphics[
height=2.3329in,
width=2.9408in
]%
{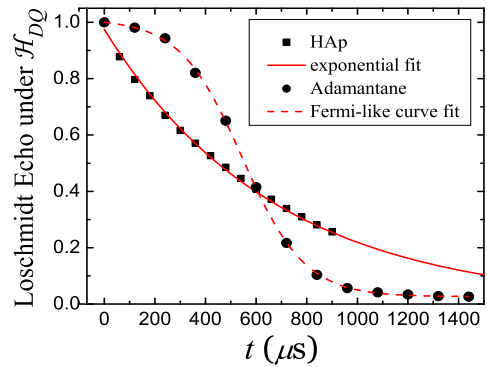}%
\caption{(Color online) Loschmidt echo experiment based on $\mathcal{H}_{DQ}$
and $-\mathcal{H}_{DQ}$ using the sequence shown in Fig.\ref{Fig_pulsos}(c)
with $\phi=0$ in HAp (squares) and adamantane (circles).}%
\label{Fig_experiments_echoe}%
\end{center}
\end{figure}

In order to have a measure of the global decoherence time of the
system\textbf{ }under $\mathcal{H}_{DQ}$, we used the eight-pulse sequence
shown in Fig. \ref{Fig_pulsos}(c). Having minimized possible experimental
artifacts, we expect to have a decoherence that reflects the properties of the
sample itself (topology of the coupling network, defects, etc.). Following
this idea, we compare the behaviors of HAp and adamantane measuring a
Loschmidt echo, that is, generating $\mathcal{H}_{DQ}$ and then $-\mathcal{H}%
_{DQ}$. The decays of both systems are displayed in Fig.
\ref{Fig_experiments_echoe}. The difference in the functional form of decay is
remarkable. While a simple exponential with characteristic time $\tau
_{\varphi}=\left(  770\pm50\right)  \ \mu s$ holds for HAp, a Fermi-type curve
$M(t)\propto1/\left[  1+\exp\left(  (t-t_{c})/\tau_{\varphi}\right)  \right]
$ with $t_{c}=(545\pm2)\ \mu s$ and $\tau_{\varphi}=(123\pm2)\ \mu s$,
provides the best fit for adamantane.%

It should be noticed that in adamantane, coherences of very high order are
generated quite rapidly. Indeed, coherences of orders $M>100$ are well defined
after $0.5$ $ms$ \cite{KS04, SLAC08}, indicating the huge portion of the
Hilbert space explored through $\mathcal{H}_{DQ}$ in this system. As shown by
the Fermi-type curve, the coherence of such highly interacting system is not
sustained beyond a critical time $t_{c}$ where a sort of \textquotedblleft
catastrophe\textquotedblright\ seems to occur. A similar behavior is observed
in simulations of highly interacting systems, either fermions or bosons, whose
coherence also decays following a Fermi-type curve \cite{MH06, MH08}. In those
works, a\emph{ }self-consistent approximation allows one to see this critical
stage as the triggering of a nonlinear loop.

In contrast with adamantane, the decoherence of HAp, as seen from the
Loschmidt echo, occurs smoothly following an exponential law. This sort of
decay has been seen in chaotic one-body systems in semiclassical states where
the perturbation effects are limited \cite{CPJ04, GPSZ06, JP08}. Hence, this
decay is consistent with the restricted dynamics imposed by the low
connectivity of a 1d system. Furthermore, as the dominant dynamics is that of
the non-interacting fermions, the residual interactions and the experimental
imperfections define the \textquotedblleft environment\textquotedblright\ that
produces the exponential decoherence.

Although the observed decoherence rate seems to be somewhat fast to enable a
straightforward quantum information application, the exponential decay of the
dynamics of the 1d system may be easier to manipulate than the dynamics of the
3d system. On the other hand, the 3d system presents a short-time behavior
that could be nicely exploited to implement quantum operations, because the
coherence is lost at a very low rate. Further experimental designs are
necessary to confirm the origin of these different functional forms, and to
quantify the factors determining the respective characteristic decay times
$\tau_{\varphi}$ in HAp and $t_{c}$ and $\tau_{\varphi}$ in adamantane.

\section{Conclusion}

We have shown that the $M$-QC intensities under a double-quantum Hamiltonian
in HAp behave as effective one-body dynamics. This has been observed through
several experiments where the evolutions of the intensities of the $2$-QC and
$4$-QC were studied under the action of $\mathcal{H}_{DQ}$. These results were
contrasted with the many-body dynamics induced by $\mathcal{H}_{XX}$. No
coherence orders above 2 appear under $\mathcal{H}_{DQ}$, while they do appear
under $\mathcal{H}_{XX}$. In both cases, the dynamics remains mainly
one-dimensional as the natural anisotropy of HAp is enhanced by the quantum
Zeno effect.

The global decoherence of HAp under $\mathcal{H}_{DQ}$ was compared with that
in adamantane, a regular 3d system, whose genuine many-body dynamics is
manifested by the rapid excitation of very high orders of coherence. The
coherence decays in both systems follow completely different functional forms.

In summary, we have addressed two main points:

(1) We confirmed the mapping of a nearest-neighbor one-dimensional spin system
under a double-quantum interaction to a non interacting fermion system. This
mapping was tested through one of its main consequences: the non excitability
of $4$-QC under $\mathcal{H}_{DQ}$.

(2) We evaluated the decoherence through a Loschmidt echo experiment based on
a double-quantum Hamiltonian. The restricted dynamics induced by the low
connectivity space leads to the appearance of a smooth exponential
decoherence, while the dynamics in a high connectivity space shows a sudden
drop in coherence.

These results indicate that, in spite of residual interactions, HAp can be
used as a \textquotedblleft quantum simulator\textquotedblright\ for non
interacting fermion dynamics.

\begin{acknowledgments}
The authors thanks R. H. Acosta, Y. Garro Linck and A. K. Chattah for fruitful
experimental discussions, as well as L. E. F. Foa Torres and F. Pastawski for
suggestions. This work was made possible through the financial support from
CONICET, ANPCyT and SeCyT-UNC. PRL and HMP acknowledge the hospitality of the
Institute for Materials Science TU-Dresden and the MPI-PKS.
\end{acknowledgments}

\bibliographystyle{apsrev}
\bibliography{ele_st_last}

\end{document}